\documentclass[prb,aps,amsmath,amssymb,twocolumn]{revtex4-2}
\usepackage{graphicx,amsmath,amssymb,color,xcolor}
\usepackage{setspace}
\usepackage[utf8]{inputenc}
\usepackage{ulem}
\usepackage{nicefrac}
\usepackage{multirow, makecell, ragged2e}
\usepackage[titletoc,title]{appendix}
\usepackage[colorlinks,bookmarks=true,citecolor=blue,linkcolor=red,urlcolor=blue]{hyperref}
\usepackage{cancel}
\usepackage{bm}
\usepackage{orcidlink}

\newcommand{\be}{\begin{equation}}
\newcommand{\ee}{\end{equation}}

\newcommand{\ba}{\begin{eqnarray}}
\newcommand{\ea}{\end{eqnarray}}

\newcommand{\LLL}{\text{P}_{\text{LLL}}}

\newcommand{\conj}[1]{\bar{#1}}

\def\beq{\begin{eqnarray}}
\def\eeq{\end{eqnarray}}

\newcommand{\abs}[1]{\vert #1 \vert}

\newcommand{\ket}[1]{\vert #1 \rangle}
\newcommand{\bra}[1]{\langle #1 \vert}
\newcommand{\overlap}[2]{\langle #1 \vert #2 \rangle}

\newcommand{\Mod}[2]{#1\text{ mod }#2}
\newcommand{\AB}{Aharonov–Bohm~}
\begin{document}

\title{Proposal for bulk measurement of braid statistics in fractional quantum Hall effect}
\author{Mytraya Gattu,$^1$ G.J. Sreejith,$^2$\orcidlink{0000-0002-2068-1670} and J. K. Jain$^1$~\orcidlink{000-0003-0082-5881}}
\affiliation{$^1$Department of Physics, 104 Davey Lab, Pennsylvania State University, University Park, Pennsylvania 16802,USA}
\affiliation{$^2$Indian Institute of Science Education and Research, Pune, India 411008}
\date{\today}
\begin{abstract}
The quasiparticles (QPs) or quasiholes (QHs) of fractional quantum Hall states have been predicted to obey fractional braid statistics, which refers to the Berry phase (in addition to the usual Aharonov-Bohm phase) associated with an exchange of two QPs or two QHs, or equivalently, to half of the phase associated with a QP/QH going around another.  Certain phase slips in interference experiments in the fractional quantum Hall regime have been attributed to fractional braid statistics, where the interference probes the Berry phase associated with a closed path which has segments along the edges of the sample as well as through the bulk (where tunneling occurs).  Noting that QPs / QHs with sharply quantized fractional charge and fractional statistics do not exist at the edge of a fractional quantum Hall state due to the absence of a gap there, we provide arguments that the existence of composite fermions at the edge is sufficient for understanding the primary experimental observations; composite fermions are known to occur in compressible states without a gap. We further propose that transport through a closed {\it tunneling} loop contained entirely in the bulk can, in principle, allow measurement of the braid statistics in a way that the braiding object explicitly has a fractionally quantized charge over the entire loop. Optimal parameters for this experimental geometry are determined from quantitative calculations.
\end{abstract}
\maketitle

\section{Background}

Laughlin demonstrated the existence of quasiparticles (QPs) and quasiholes (QHs) with fractional local charge $qe$ in the fractional quantum Hall (FQH) effect by the trick of adiabatic insertion of a point flux quantum~\cite{Laughlin83}. (The ``local charge" is defined as the charge excess or deficiency in a finite area relative to the ground state.) Subsequently, Halperin proposed that these excitations also obey fractional braid statistics~\cite{Halperin84}, i.e., the Berry phase factor associated with a closed loop of a QP  / QH acquires an additional factor $e^{i\theta}$ when another QP/QH is added in the interior of the loop, where $\theta$ depends on the background FQH state but is independent of the size or the shape of the closed loop. The existence and the allowed values of both the fractional charge and fractional statistics can be deduced based on general principles by assuming incompressibility at a fractional filling factor~\cite{Su86} and does not require a detailed microscopic understanding of either the ground state or the excitations. For the standard states at $\nu=n/(2pn\pm 1)$, which will be our focus below, $q$ and $\theta$ take the simplest allowed values allowed by general considerations:  $|q|=1/(2pn\pm 1)$ ($q$ is defined in units of the electron charge $|e|$) and $|\theta|=2\pi \times 2p/(2pn\pm 1)$~\cite{Jain07}. Explicit calculation based on accurate model wave functions of the composite fermion (CF) theory have confirmed these values~\cite{Arovas84,Kjonsberg99,Kjonsberg99b,Jeon03b,Jeon04,Tserkovnyak03,Nardin23,Trung23,Bose24}. 

\begin{figure}
\includegraphics[width=0.7\columnwidth]{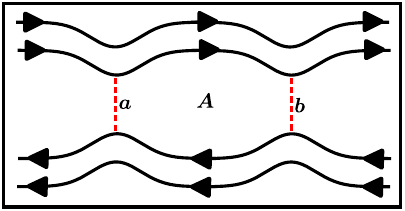}
\caption{This schematic diagram shows the geometry used in interference experiments, where a closed loop enclosing an area $A$ involves the innermost edge channel (solid black lines) and two tunneling segments a and b (dashed red lines). Note that the drawing is not to scale. 
\label{Fig1}}
\end{figure}

A possible method for measuring the statistics is to consider the geometry given in Fig.~\ref{Fig1}a ~\cite{Chamon97} and imagine a QP impinging from the upper left edge. It can end up at the lower left edge either by tunneling across the constriction a or by tunneling across the constriction b. An interference between these paths encodes information regarding the Berry phase associated with the loop enclosing area $A$ in Fig.~\ref{Fig1}.  The experiment in Ref.~\cite{Nakamura20} has implemented this geometry and made two key observations in a range of parameters near $\nu=1/3$: (i) the conductance as a function of magnetic flux through the loop has a period of $3\phi_0$, where $\phi_0=h/e$ is  the flux quantum, and (ii) there are occasional phase slips of $-2\pi/3$. These observations are consistent with a fractionally charged QP moving along the edge, with the phase associated with its closed loop undergoing a fractional phase slip when another QP is added in the bulk due to their fractional braid statistics.  More specifically, the period of $3\phi_0$ 
is interpreted in terms of a charge $|q|=1/3$; and the phase slips are thought to occur due to the addition or removal of a QP in the interior and is consistent with the statistics parameter $\theta=-2\pi/3$. These observations have been confirmed in Refs.~\cite{Werkmeister24,Samuelson24} and extended to $\nu=2/5$ in Ref~\cite{Nakamura23}.

The dynamics of QPs and QHs in the bulk has been modeled in terms of fractionally charged anyons~\cite{Halperin84,Kivelson90,Jain93,Hansson96,Hansson09,Rosenow20} which may capture certain topological features. However, QPs or QHs with sharply quantized $q$ or $\theta$ do not exist at the edge of a FQH system because of the absence of a gap at the edge. This has been seen in explicit evaluations of these quantities using accurate trial wave functions~\cite{Arovas84,Kjonsberg99,Kjonsberg99b,Jeon03b,Jeon04,Tserkovnyak03,Nardin23,Trung23,Bose24}, which find that $q$ and $\theta$ are well defined when the QPs / QHs are in the bulk and sufficiently far separated (there is a correction when their density profiles overlap), but cease to be well defined when they are near the edge of the sample. It appears plausible that one should be able to create a lump of arbitrary local charge at the edge of a FQH system. The subtleties associated with the precise definition of charge in an ordinary one-dimensional chiral Luttinger liquid have been  analyzed in Ref.~\cite{Leinaas09}, which shows that excitations of arbitrary local charges can be created in this liquid. How do we then understand the experimental results?

Chamon {\it et al.}~\cite{Chamon97} had formulated a theory of the interference experiments in 1997 using the Luttinger liquid approach for $\nu=1/(2p+1)$ states with a single chiral edge mode (assuming the absence of edge reconstruction), which is meant to provide an effective description of the problem in the long-wavelength low-energy limit. This approach does not assume fractionally charged quasiparticles at the edge, but assumes that the objects that tunnel at the constrictions are fractionally charged, which would be expected to be the case when the constriction is sufficiently wide to comfortably enclose a QP/QH, i.e., when the tunnel amplitude is sufficiently small. The predictions from this approach are consistent with the experiment. More recently, Feldman and Halperin~\cite{Feldman22} demonstrated, also using the Luttinger liquid approach, that the finite width of the edge states (which renders the enclosed area ill-defined) and the presence of  localized anyons in the vicinity of the edge states do not alter the result in the low-temperature limit for situations when a single chiral edge is present.

We analyze the experiment in terms of CFs. The fundamental property of CFs is that each CF sees an even number of vortices at every other CF, which implies that the CFs have a nontrivial (but integral) braiding statistics. As discussed earlier (and reviewed below), this property leads to fractional charge and fractional braid statistics for the QPs and QHs in the bulk of an incompressible state.  However, unlike fractionally charged QPs / QHs, CFs can be well defined even in a compressible state, such as those in the 1/2 CF Fermi liquid or at the edge of a FQH state. We argue in this article that an understanding of the interference experiments does not require sharply quantized fractional charge at the edge, but only that CFs be well defined there.

We further ask whether an in-principle realizable geometry can implement braiding such that the braiding object explicitly has a fractionally quantized charge over the entire loop. We propose that this may be accomplished by 
considering a closed {\it tunneling} loop of a QP/QH contained entirely in the bulk, demonstrating that in appropriate parameter regimes, a tunneling loop produces the same Berry phase as a continuous loop does. Such a tunneling loop can be created by a careful placement of impurities that provide preferred locations for a QP/QH injected into the bulk. Coupling of this loop to the two edges and studying the magnetic field dependence of the inter-edge conductance for transport through this loop can provide a direct measurement of fractional charge as well as fractional statistics of the QPs / QHs. We determine the optimal parameters for such a measurement by detailed calculation.

The plan for the rest of the article is as follows. In Sec.~\ref{sec:What} we show that the experimental observations can be understood as a measure of the relative braid statistics of a CF and a QP/QH.  We then propose in Sec.~\ref{sec:Proposal} an alternative geometry for measuring the relative statistics of QPs or QHs in an unambiguous manner. The idea is to create a closed tunneling loop in the bulk of the system, away from the edges, by placing a string of impurities . The paper is concluded in Sec.~\ref{sec:Discussion}. 

\section{What do interference experiments measure?}  
\label{sec:What}
 
The understanding of the FQH states at $\nu=n/(2pn\pm 1)$ as the $\nu^*=n$ integer quantum Hall (IQH) states of weakly interacting CFs will form the basis for our discussion below.  We begin here with a brief review of the aspects of the CF theory that are necessary for the present purposes~\cite{Jain89,Jain07,Halperin20}. In what follows, the Berry phase factor associated with a closed loop will be defined as $e^{\imath \gamma}$.

\subsection{Integral braid statistics of the CFs}

A CF is the bound state of an electron and an even number ($2p$) of vortices~\cite{Jain89,Jain07}. It is often modeled as the bound state of an electron and $2p$ flux quanta, where a single flux quantum is defined as $\phi_0=h/e$. As in Wilczek's model of anyons~\cite{Wilczek82}, because of the attached flux, a closed counterclockwise loop of a CF around another CF produces a phase factor $e^{i\theta_{\rm CF-CF}}$ with 
\begin{equation}\label{eq:berry-phase-CF-CF}
\theta_{\rm CF-CF}=4p\pi.
\end{equation}
(Note that this phase is in addition to the fermionic exchange statistics of the electrons, which is incorporated through an explicit anti-symmetrization of the wave function.) Although not usually stated as such, the CFs obey ``integral braid statistics" (as $\theta_{\rm CF-CF}$ is $2\pi$ times an integer), which is what makes them topologically distinct from electrons.
When a CF traverses a loop enclosing an area $A$, it acquires a Berry phase
\begin{equation}\label{eq:berry-phase-mean-field}
\gamma=-2\pi {BA\over \phi_0} +  \theta_{\rm CF-CF} N_{\rm enc}
\end{equation}
where $N_{\rm enc}$ is the number of CFs (or electrons) enclosed inside the  loop. The first term on the right hand is the AB phase of an electron of charge $-e$ going around a loop of area $A$ in a magnetic field. The second term arises from the braid statistics of the CFs.

The best known consequence of the CFs' braid statistics is that the CFs effectively experience a much reduced magnetic field $B^*$. This follows from a mean field approximation in which one replaces $N_{\rm enc}$ by its average value $N_{\rm enc}=\rho A$, where $\rho$ is the electron or the CF density, and interprets the Berry phase $\gamma$ as the \AB (AB) phase $-2\pi B^*A/\phi_0$ due to a reduced magnetic field $B^*=B-2p\rho \phi_0$. CFs form their own Landau levels (LLs), called $\Lambda$Ls ($\Lambda$Ls), in the reduced magnetic field, and fill $\nu^*$ of them, which is related to the electron filling factor $\nu$ by the relation $\nu=\nu^*/(2p\nu^*\pm 1)$, where the $+(-)$ sign applies when $B^*$ is parallel (antiparallel) to $B$. The reduced magnetic field $B^*$, the CF filling factor $\nu^*$, and $\Lambda$Ls are all a direct consequence of the CFs' braid statistics. Experimental confirmations of $B^*$, $\nu^*$ and $\Lambda$Ls, and hence of the braid statistics of the CFs, are too numerous to list in their entirety here, but some of the prominent ones are~\cite{Jain07,Halperin20}: FQH effect (FQHE) at the Jain sequences $\nu=n/(2pn\pm 1)$~\cite{Jain89,Jain07}; the CF Fermi sea at $\nu=1/2$~\cite{Halperin93,Halperin20b,Shayegan20}; quantum (or Shubnikov-de Haas) oscillations around $\nu=1/2$~\cite{Du94,Leadley94}; the CF cyclotron orbits~\cite{Willett93,Kang93,Goldman94}. All of the experimental successes of the CF theory ultimately arise from their integral braid statistics. As discussed in the next subsection, the fractional braid statistics of the QPs / QHs is also a manifestation of the CFs' integral braid statistics.

A crucial observation relevant to the present discussion is that the integral braid statistics of the CFs is valid independent of whether they belong to the ground state or are excited CFs, and whether the state is incompressible or compressible. In particular, the integral braid statistics is not associated with any sharply defined local charge, fractional or otherwise.  Indeed, many of the above mentioned phenomena refer to the compressible CF Fermi sea state~\cite{Halperin20b,Shayegan20}, where the integral braid statistics of the CFs remains well defined and well confirmed, even though the system is gapless and does not support excitations with sharply quantized local charge. (The charge $-e$ appearing in the first term in Eq.~\ref{eq:berry-phase-mean-field} is the AB charge of the CFs, determined by its coupling to the external magnetic field.) In particular, the CF description has also been found to accurately capture finite systems with an edge, demonstrating that CFs remain well defined at the edge of a FQH state~\cite{Jain95,Jeon04b,Jeon07,Jain07}.

\subsection{QPs / QHs and their braid statistics}

\begin{figure}[t]
\includegraphics[width=0.45\columnwidth]{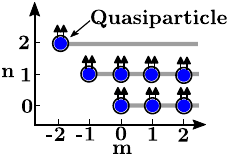}
\includegraphics[width=0.45\columnwidth]{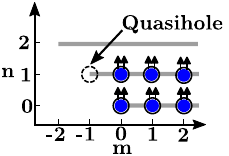}
    \caption{This figure shows schematically the QP and the QH in the lowest angular momentum orbital. The QP is a solitary CF (shown as an electron carrying two arrows) in an otherwise empty $\Lambda$L and the QH is a missing CF in an otherwise filled $\Lambda$L. The y-axis shows the $\Lambda$L index $n=0, 1, \cdots$, and the x-axis is the angular momentum index $m=-n,-n+1,\cdots$.}
    \label{fig:QPQH}
\end{figure}

In this subsection, we review how the fractional braid statistics of the QPs and QHs also arises as a direct consequence of the CFs' integral braid statistics. Let us begin by recalling the definition of a QP/QH.  

The incompressible state at $\nu=n/(2pn+ 1)$ is $n$ filled $\Lambda$Ls of CFs. Consider now the many-particle state with the lowest $n$ $\Lambda$Ls fully occupied and a single CF in the $(n+1)^{\rm th}$ $\Lambda$L. It is convenient to view this many-particle state as a {\it single} QP, relative to the ``vacuum" state of $n$ filled $\Lambda$L state. Similarly, the state with the lowest $n$ $\Lambda$Ls fully occupied except for a missing CF in the $n^{\rm th}$ $\Lambda$L is viewed as a {\it single} QH. See Fig.~\ref{fig:QPQH} for a pictorial representation. Remarkably, the CF theory provides an almost perfect quantitative account of the actual QPs and QHs (see, for example, quantitative comparisons with exact diagonalization studies in Ref.~\cite{Gattu24}).  

The local charge can be derived as follows. Let us create a QP by adding a CF, i.e. an electron carrying $2p$ vortices, to an incompressible state. For this purpose, we perform an adiabatic insertion of $2p$ vortices at some location followed by the addition of an electron. Noting that each vortex produces a local charge $\nu$ (in units of $|e|$) relative to the uniform charge incompressible state, the local charge of a QP is given by $q= -1+2p\nu = -1/(2pn+ 1)$. (We added a unit charge overall; the rest of its charge is pushed to the boundary.) The QH, which is obtained by removing a CF, has a charge is $q= +1/(2pn+1)$, which follows from the fact that a QP-QH pair has no net local charge. 

To see how fractional braid statistics arises, let us ask: What is the change in the Berry phase of a closed path of a CF enclosing an area $A$ when a QP or QH is inserted inside it? Following Eq.~\ref{eq:berry-phase-mean-field}, the answer  is:
\begin{equation}\label{eq:CF-qp-qh-statistics}
\theta=\Delta \gamma = 4p\pi \Delta N_{\rm enc}=\pm {4p\pi \over 2pn +1}, 
\end{equation}
where the $+$ ($-$) sign is for the insertion of a QP (QH). 
This answer is valid independent of whether CF going around in the loop is an excited CF (i.e. a QP) or a CF in the ground state. We have thus determined four statistical angles:
\begin{equation} \label{eq:CF-qp-qp-qp-statistics}
\theta_{\rm CF-QP}={4p\pi \over 2pn +1}, \;\; \theta_{\rm QP-QP}={4p\pi \over 2pn +1},
\end{equation}
\begin{equation}\label{eq:CF-qh-qp-qh-statistics}
\theta_{\rm CF-QH}=-{4p\pi \over 2pn +1}, \;\; \theta_{\rm QP-QH}=-{4p\pi \over 2pn +1}.
\end{equation}
Here, the subscript ``CF" specializes to a CF in the ground state.
We stress that even a CF in the ground state has a well defined braid statistics relative to a QP or a QH.  Finally, because a QP-QH pair is a boson, it follows that 
\begin{equation}\label{eq:qh-qh-statistics}
\theta_{\rm QH-QH}={4p\pi \over 2pn +1}.
\end{equation}
The fractional braid statistics thus results from the integral braid statistics of the CFs combined with the fractional local charge of a QP or a QH. Alternatively, one may note that a QP/QH carries a fractional flux of magnitude $2p/(2pn\pm 1)$ along with it. (Unlike the standard statistics which arises as a consequence of indistinguishability, the braid statistics is defined even for distinguishable particles, such as a QP going around a QH, in which case it is referred to as ``relative" statistics.)

Explicit calculations using the wave functions of the CF theory have confirmed these results. 
One can calculate the charge excess or deficit (i.e. the local charge) by integrating the density obtained from the microscopic wave functions, and the fractional statistics by a Berry phase calculations~\cite{Arovas84,Kjonsberg99,Kjonsberg99b,Jeon03b,Jeon04,Tserkovnyak03,Nardin23,Trung23,Bose24}.
The microscopic calculations also bring out the limitations of the concepts of fractional charge and statistics. The QPs and QHs satisfy sharp fractional braid statistics provided that they are non-overlapping and far from the edges. These calculations (and also the ones below) show that the fractional charge and statistics do not remain well defined when the QP/QH approaches the edge of the system.  The description in terms of the CFs continues to be valid, however.

We remark that while $B^*$ is an order-$N$ (i.e. thermodynamic) effect, the fractional braid statistics of the QPs or QHs appears as a rather subtle, order-one change in the the Berry phase. That is one of the reasons why the latter has proved rather difficult to detect in experiments.

\subsection{Interference experiments}

As noted above, excitations with sharp fractional charge cannot be defined at the edge of a FQHE state, given that there is no gap at the edge. In trial wave functions, when QPs and QHs which have well defined charges in the bulk are brought close to the edge, their local charges no longer remain quantized. We now argue that the fractional phase slips in the interference experiments can be understood so long as CFs are well defined at the edge.

Let us imagine a closed orbit at the Fermi energy that coincides with the edges and passes through the quantum point contacts at the positions of maximum tunneling, as shown in Fig.~\ref{Fig1}. We define the area of the region inside orbit to be $A$. A maximum in backscattering is obtained when this orbit coincides with a quantized orbit of a CF.  Note that a CF moving along the edge is a part of the ground state, but when the CF tunnels from one edge to the other through the point contact, it tunnels as a fractionally charged QP.  However, we can treat it as a CF over the entire loop for the purposes of fractional statistics because $\theta_{\rm QP-QP}=\theta_{\rm CF-QP}$ and $\theta_{\rm QP-QH}=\theta_{\rm CF-QH}$.

Let us now consider a FQH state at filling factor $\nu=n/(2pn\pm 1)$.  The region of interest can respond in several ways as an external parameter such as the magnetic field or the gate voltage is varied.

(i) The region evolves such that the number of QPs or QHs inside the loop does not change. In this case the number of the CFs enclosed by the loop ($N_{\rm enc}$) changes continuously, i.e. the CFs flow into or out of the loop through the open quantum point contacts as a part of the ground state. (Recall that in the experimental geometry, the quantum point contacts are $\sim$97\% open.) The area $A$ enclosed by the loop may also change continuously. The Berry phase associated with the closed CF path is given by Eq.~\ref{eq:berry-phase-mean-field} with $N_{\rm enc}=\rho A=B A \nu/\phi_0$ where $\rho$ is the average density and $\nu$ the average filling:
\begin{equation}\label{eq:berry-phase-fqhe}
\gamma=-2\pi {BA\over \phi_0} + 4p\pi {B A \nu \over \phi_0}
\end{equation}
From the semiclassical Bohr-Sommerfeld quantization condition, quantized orbits of the CF are obtained when the Berry phase is $2\pi$ times an integer, which implies a period of $\Delta \gamma=2\pi$ for two successive quantization conditions. This yields a period 
\begin{equation}\label{eq:fqhe-period}
\Delta (BA)={1\over 2p\nu -1} \phi_0 
\end{equation}
We note that this period changes continuously with the average filling factor $\nu$ inside the area $A$. In particular, for $\nu=n/(2n\pm 1)$ this gives a period of $(2n\pm 1)\phi_0$. 

(ii) In the second process, the number of QPs/QHs inside the loop changes, but without inducing any change at the edge, i.e. the CF at the Fermi energy has exactly the same closed loop as that without the additional QPs/QHs. (When there is an interaction between the additional charge and the edge, the analysis becomes complicated. One of the accomplishments of Ref.~\cite{Nakamura20} was to minimize this interaction with the help of screening layers.) The change in the Berry phase is 
\begin{eqnarray}
\Delta \theta &= &\Delta N_{\rm QP}\theta_{\rm CF-QP}+\Delta N_{\rm QH}\theta_{\rm CF-QH}\nonumber \\
&=&\Delta (N_{\rm QP}-N_{\rm QH}){4p \pi\over 2pn \pm 1}
\end{eqnarray}
For $\nu=1/3$ ($n=p=1$), this gives a phase slip of $4\pi /3$ for the addition of a QP ($\Delta N_{\rm QP}=1, \Delta N_{\rm QH}=0$), which is equivalent to $\Mod{-2\pi/3}{2\pi}$. For the addition of a QH ($\Delta N_{\rm QP}=0, \Delta N_{\rm QH}=1$) at $\nu=1/3$ we get a phase slip of $+2\pi/3$.

(iii) Finally, it is possible that when the number of QPs/QHs inside the loop changes, it alters the electrostatic potential at the edges thereby  causing a change in the energies and the occupations of the CF orbitals at the edge. This complicates the analysis~\cite{Halperin11,Feldman21} and will not be discussed in this article.

The above argument is valid for any filling factor and is also independent of whether the edge is reconstructed or not. However, it is expected that the condition of ``no change at the edge when a QP/QH is added in the interior" is easiest to satisfy when there is a single chiral edge mode, as is the case at $\nu=1/3$ in the absence of edge reconstruction. When the edge consists of several modes, as is the case for $\nu=n/(2pn\pm 1)$ with $n>1$ (Fig.~\ref{Fig1}b), or when the edge is reconstructed, the addition of a QP/QH in the bulk may cause charge rearrangements between the edge modes of different $\Lambda$Ls, thus making the FQH edge more susceptible to perturbation.

\section{A bulk proposal for measuring QP/QH braid statistics}
\label{sec:Proposal}

\subsection{Intuitive Idea}
\label{subsec:Proposal-Intuitive-Idea}

\begin{figure}
\includegraphics[width=1.0\columnwidth]{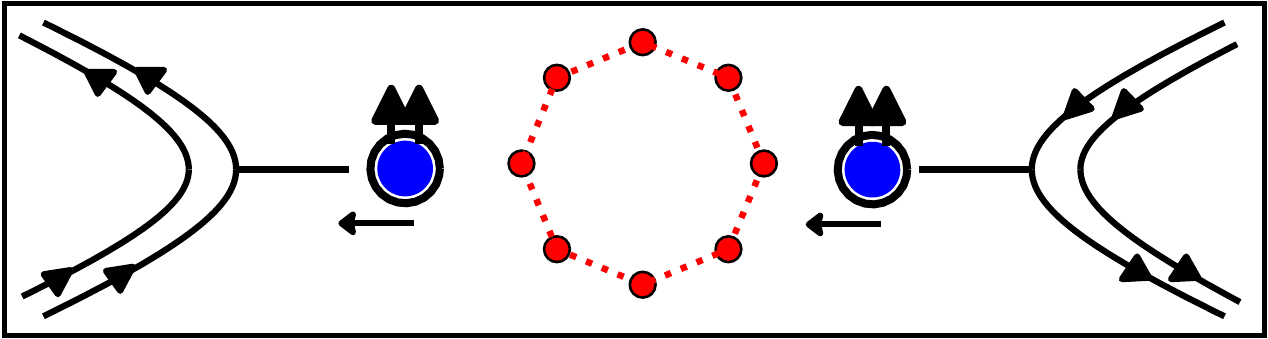}

\vspace{1cm}
\includegraphics[width=\columnwidth]{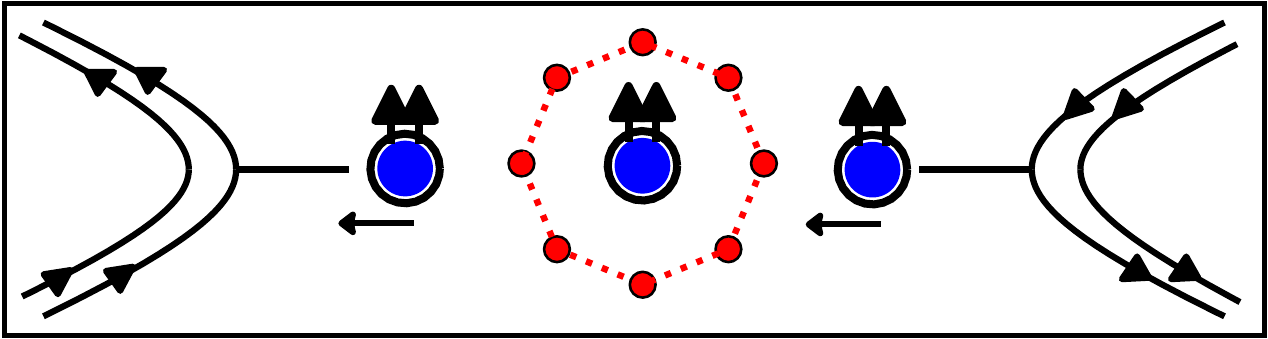}
    \caption{This figure shows a schematic setup to measure fractional charge and fractional statistics of quasiparticles (QPs) and quasiholes (QHs) using an interfering path defined in the bulk of the FQH liquid. Impurities (red dots) are placed to produce a tunneling loop (dotted red line), which is tunnel-coupled to the innermost of the edge states (solid black lines) on two sides. 
    The inter-edge tunnel conductance will exhibit maxima (minima) when the phases acquired by the QP/QH along the upper and lower paths interfere constructively (destructively), and will exhibit periodic oscillations as a function of $B$. The period of the oscillations is related to the charge of the QP/QH, and the shift in the oscillations when a QP/QH is added in the interior of the loop (lower panel) will reveal fractional statistics. It is assumed that the nearest neighbor tunneling dominates, and that the tunneling loop completely encloses the central QP/QH in the lower panel.}

    \label{fig:setup}
\end{figure}

We next ask if is in principle possible to measure the braid statistics such that the object going around the loop has a fractional charge over the entire loop. We suggest that this can be accomplished by placing a sequence of impurities in such a way as to create a closed tunneling loop in the bulk of the sample and studying the transport of a QP/QH across it, as shown in Fig.~\eqref{fig:setup}. Note that the impurities should provide states in the gap (in between the $\Lambda$Ls) so that they do not excite a QP or QH from the ground state, but only provide preferred positions for a QP/QH injected from the edge. Furthermore, we will assume that the impurity potential is sufficiently weak that any distortion in the wave function of either the ground state or the QP/QH is negligible.

The conductance for transport through the loop will be determined by the difference in the phases acquired by the QP/QH passing through the upper and lower paths, which is also the phase associated with a complete loop of a QP/QH, and will oscillate as a function of the magnetic flux enclosed by the loop. A phase shift due to the insertion of another QP/QH in the interior of the loop will be a measure of the fractional braid statistics. This we believe will constitute an unambiguous measure of the braid statistics of the QPs / QHs because the tunneling path is fully contained in the bulk, making it highly probable that the object that is hopping along the impurities has a fractional charge.  

In what follows, we perform explicit calculations for the phase associated with a tunneling loop of a QP/QH. A condition for the measurement of fractional statistics in this geometry is that the nearest neighbor impurity sites are separated by distances larger than the QP/QH sizes, so that a single loop with a well defined area is relevant. When there is appreciable tunneling into impurities beyond the nearest neighbors, multiple closed loops contribute, which introduces an uncertainty in both the Berry phase and the area enclosed. Additionally, the braid statistics becomes well defined only when the loop fully encloses the central QP/QH.

The Berry phase for a ballistic motion along a continuous loop has been determined earlier~\cite{Kjonsberg99,Kjonsberg99b,Jeon03b,Jeon04}. One can calculate the Berry phase by discretizing this path, but that is very different from the geometry considered here. 
The discretized version of the continuous ballistic path involves successive positions that are ideally infinitesimally close to one another, and tunneling beyond the nearest neighbors is not a meaningful issue. In contrast, in the geometry considered here, the successive positions of a QP/QH must lie sufficiently far so that a single loop dominates.

\subsection{Theory background: wave functions}
In this section, we introduce the Jain CF wave functions describing the incompressible ground states and their QPs and QHs for the FQH states at $\nu =n/(2pn+1)$. We will assume the disk geometry, and work with the symmetric gauge.

The incompressible FQH states at electron filling factors $\nu = n/(2pn+1)$ correspond to incompressible IQH states of the CFs at the CF filling $\nu^{\star} = n$. The wave function of the incompressible state of $N$ electrons at filling $\nu = n/(2pn+1)$  is written as the state of $N$ CFs filling $n$ $\Lambda$Ls as~\cite{Jain07}
\begin{equation}\label{eq:CF-ground-state}
    \begin{split}
        \Psi_{\frac{n}{2pn+1}}(z_{1}, \dots, z_{N}) &= \LLL \phi_{n}(z_{1}, \dots, z_{N}; B^{\star})\\
        &\times [\phi_{1}(z_{1}, \dots, z_{N})]^{2p}\\
        \phi_{1}(z_{1}, \dots, z_{N}) &= \prod_{i<j=1}^{N}(z_{i}-z_{j})e^{-\sum_{k=1}^{N}\frac{\abs{z_k}^{2}}{4\ell_{1}^{2}}}\\
    \end{split}
\end{equation}
Here, $z_{i} = x_{i} - \imath y_{i}$ are the complex coordinates of the $i^{\rm th}$ electron; $\LLL$ is the projection operator into the lowest LL; the state $\phi_{n}$ is a Slater determinant of $N$ electrons filling $n$ LLs at magnetic field $B^{\star} = B/(2pn+1)$; and the Jastrow factor $\phi_{1}^{2p}$ attaches $2p$ vortices to each electron with $\ell_{1} \equiv \ell \sqrt{(2pn+1)/n}$.
The single particle wave functions in the $n^{\rm th}$ LL are given by
\begin{equation}\label{eq:landau-level}
\begin{split}
            \eta_{n, m}(z; B^{\star}) &= \frac{(-1)^{n}}{\sqrt{2\pi}}\sqrt{\frac{n!}{2^{m}(m+n)!}} \\
        &\times \left(\frac{z}{l^{\star}}\right)^{m}L_{n}^{m}(\frac{\abs{z}^{2}}{2\ell^{\star}{}^{2}})e^{-\frac{\abs{z}^{2}}{4\ell^{\star}{}^{2}}}.
\end{split}
\end{equation}
Here, $n=0,1,2, \cdots$ specifies the LL index (note that in the phrase ``$n^{\rm th}$ LL / $\Lambda$L," $n$ refers to the LL / $\Lambda$L index); the angular momentum $L_{z}$ can have values $m=-n, -n+1, \dots$; and we define $\ell^{\star}=\sqrt{\hbar c/eB^{\star}}=\ell\sqrt{2np+1}$. 
The Slater determinant wave function for the ground state, $\phi_{n}$, in Eq.~\eqref{eq:CF-ground-state} is given by 
\begin{widetext}
\begin{equation}\label{eq:slater-determinant}
    \phi_{n}(z_{1}, \dots, z_{N}) \equiv \frac{1}{\sqrt{N!}}\begin{vmatrix}
        \eta_{0, 0}(z_{1}; B^{\star}) & \dots & \eta_{0, 0}(z_{N}; B^{\star})\\
        \vdots & \vdots & \vdots \\
        \eta_{0, m_0}(z_{1}; B^{\star}) & \dots & \eta_{0, m_0}(z_{N}; B^{\star})\\
        \eta_{1, -1}(z_{1}; B^{\star}) & \dots & \eta_{1, -1}(z_{N}; B^{\star})\\
        \vdots & \vdots & \vdots \\
        \eta_{n-1, m_{n-1}}(z_{1}; B^{\star}) & \dots & \eta_{n-1, m_{n-1}}(z_{N}; B^{\star})\\
    \end{vmatrix},
\end{equation}
\end{widetext}
where $m_n$ is the largest angular momentum orbital occupied in the $n^{\rm th}$ LL.

We next construct an angular momentum basis for the lowest energy QPs and QHs. Since we aim to study the fractional charge and statistics, we shall work below with the unprojected wave functions which are much easier to evaluate numerically but have the same topological properties as the  projected wave functions. 

A QP is a CF in an otherwise empty $\Lambda$L. At filling $\nu = n/(2pn+1)$, the lowest energy QP state at angular momentum $L_{z}=m$ corresponds to adding a CF to the $n^{\rm th}$  $\Lambda$L in the $L_{z}=m$ orbital as follows:
\begin{equation}\label{eq:CF-qp}
\begin{split}
        \Psi^{\mathrm{QP},m}_{\frac{n}{2pn+1}}(z_{1}, \dots, z_{N}) &=  \phi_{n}^{\mathrm{QP}, m}(z_{1},\dots,z_{N}) \\
        &\times [\phi_{1}(z_{1}, \dots, z_{N})]^{2p}
\end{split}
\end{equation}
Here, $\phi_{n}^{\mathrm{QP}, m}(z_{1}, \dots, z_{N})$ is a Slater determinant of $N$ electrons filling the lowest $n$ LLs and the $L_{z}=m$ orbital in the $n^{\rm th}$ LL; $\phi_{n}^{\mathrm{QP}, m}$ is prepared at magnetic field $B^{\star}$. Similarly, a QH state is obtained by removing a CF from a filled $\Lambda$L. The lowest energy QH state at filling $\nu = n/(2pn+1)$ at angular momentum $L_{z}=-m$, corresponds to removing a CF from the $L_{z}=m$ orbital of the $(n-1)^{\rm th}$ $\Lambda$L as follows:
\begin{equation}\label{eq:CF-qh}
\begin{split}
        \Psi^{\mathrm{QH},m}_{\frac{n}{2pn+1}}(z_{1}, \dots, z_{N}) &= \phi_{n}^{\mathrm{QH}, m}(z_{1},\dots,z_{N}) \\
        &\times [\phi_{1}(z_{1}, \dots, z_{N})]^{2p}
\end{split}
\end{equation}
Here, $\phi_{n}^{\mathrm{QH}, m}(z_{1}, \dots, z_{N})$ is a Slater determinant of $N$ electrons filling the lowest $n$ LLs except for the $L_{z}=m$ orbital in the $(n-1)^{\rm th}$ LL.

\begin{figure}
\includegraphics[width=0.45\columnwidth]{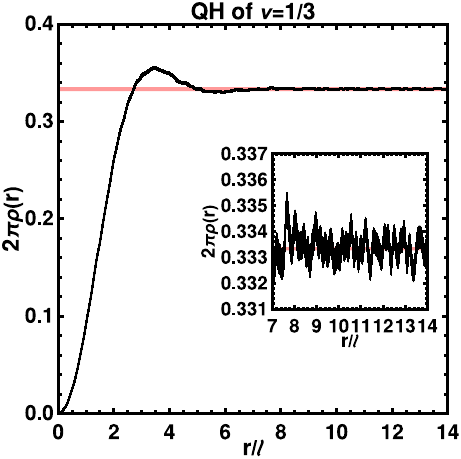}
\includegraphics[width=0.45\columnwidth]{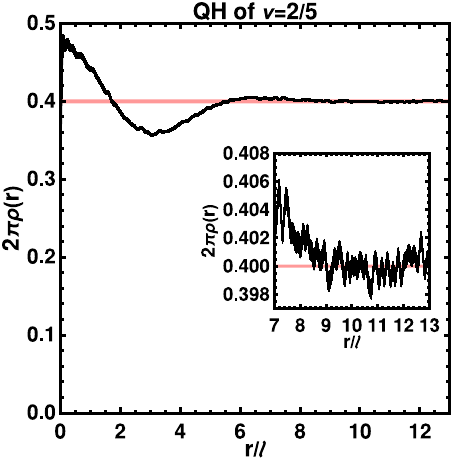}
\caption{The density profile $\rho(r)$ of the (unprojected) QH localized at the origin $r=0$ at filling $\nu = 1/3$ (left) and $\nu = 2/5$ (right). We have shown the uniform density of the FQH ground state, $\rho_{0} = \nu/(2\pi\ell^2)$, by a red line for reference.
The radius of the QH is $\sim 7 \ell \approx 4.0 \ell^{\star}$ at $\nu = 1/3$ and $\sim 11 \ell \approx 4.9 \ell^{\star}$ at $\nu = 2/5$ (as can be seen from the insets).
}
\label{fig:qh-densities}
\end{figure}

The QP (QH) of the smallest size in a given $\Lambda$L has the smallest angular momentum $m=-n$  ($m=-n+1$), as shown schematically in Fig.~\ref{fig:QPQH} (see the discussion below). We will assume below that these smallest QPs (QHs) are the relevant ones for our purposes. Their density profiles are shown in Fig.~\ref{fig:qh-densities}.

Let us imagine an infinitely large incompressible FQH state at filling $\nu = n/(2pn+1)$. This state is translationally and rotationally invariant.  The QP basis is given by \{$\Psi^{\mathrm{QP},m}$\}  ($\Psi^{\mathrm{QP},m}$ given by Eq.~\ref{eq:CF-qp}) where $m=-n, -n+1, \cdots$. These are all degenerate in the absence of any impurity potential. The degeneracy is removed when we place an impurity at a point $\omega$ described by a potential $\hat{V}(\abs{z-\omega})$ that produces a preferred location for a QP. Similarly, an appropriate impurity potential will lift the degeneracy of a QH. As stated above, we assume that the QP/QH localized by the impurity potential $\hat{V}(\abs{z-\omega})$ has the smallest size, i.e. has the smallest angular momentum relative to the point $\omega$. 

The wave functions for the QP is analogous to that of an electron in the $n^{\rm th}$ LL.  An electron in the $n^{\rm th}$ LL localized at the origin in the smallest angular momentum state is given by $\sim {\bar z}^n e^{-\abs{z}^{2}/4\ell^{\star}{}^{2}}$. For an electron localized at $\omega$, this wave function is modified to
\begin{equation}\label{eq:landau-level-coherent-state}
\begin{split}
     \overlap{z}{\omega} &= \left(\sum_{m=-n}^{\infty}\conj{\eta}_{0, m+n}(\omega; B^{\star})\eta_{n, m}(z; B^{\star})\right)\\
     &\sim \left(\frac{\conj{z}-\conj{\omega}}{\ell^{\star}}
     \right)^{n}\exp\left(\frac{\conj{\omega}z}{2\ell^{\star}{}^{2}}-\frac{\abs{z}^{2}}{4\ell^{\star}{}^{2}}-\frac{\abs{\omega}^{2}}{4\ell^{\star}{}^{2}}\right).
\end{split}
\end{equation}
This is also the smallest size electron wave packet that can be constructed in the $n^{\rm th}$ LL; this follows because an electron wave packet with angular momentum $L_{z}=m$ has size $2\sqrt{\langle \abs{\hat{z}-\omega}^{2} \rangle}=2\sqrt{2(m+2n+1)}\ell^{\star}$. Upon composite-fermionization, it is expected to give the smallest size wave packet in the $n^{\rm th}$ $\Lambda$L.

The wave function for the QP localized at $\omega$, denoted $\ket{\omega}_{\rm QP}$,  can be written as
\begin{equation}\label{eq:qp-coherent-state}
\begin{split}
     \overlap{z_{1}, \dots, z_{N}}{\omega}_{\rm QP} &\equiv \sum_{m=-n}^{\infty}\conj{\eta}_{0, m+n}(\omega; B^{\star})\phi_{n}^{\mathrm{QP}, m}(z_{1}, \dots, z_{N})\\
     &\times [\phi_{1}(z_{1}, \dots, z_{N})]^{2p}\\
     &\equiv \sum_{m=-n}^{\infty}\conj{\eta}_{0, m+n}(\omega; B^{\star})\Psi^{\mathrm{QP}, m}_{\frac{n}{2pn+1}}(z_{1}, \dots, z_{N})
\end{split}
\end{equation}
For convenience, we refer to the orthogonal states containing a CF in the $L_{z} = m-n$ orbital in the $n^{\rm th}$ $\Lambda$L i.e. ${\Psi}^{\mathrm{QP}, m}_{n/(2pn+1)}(z_{1}, \dots, z_{N})$ as $\ket{m}_{\mathrm{QP}}$. 
The state containing two QPs localized at $\omega_{1}$ and $\omega_{2}$ is a simple extension of the above: 
\begin{equation}\label{eq:qp-qp-coherent-state}
\begin{split}
     &\overlap{z_{1}, \dots, z_{N}}{\omega_{1}, \omega_{2}}_{\mathrm{QP}} \\
     &\equiv \left[\sum_{m_{1}, m_{2}=-n}^{\infty}\conj{\eta}_{0, m_{1}+n}(\omega_{1}; B^{\star})\conj{\eta}_{0, m_{2}+n}(\omega_{2}; B^{\star})\right.\\
     &\left. \times \phi_{n}^{{\mathrm{QP}}, m_{1}, m_{2}}(z_{1}, \dots, z_{N})\right]\times[\phi_{1}(z_{1}, \dots, z_{N})]^{2p}\\
     &\equiv \sum_{m_{1} < m_{2} = -n}^{\infty}\begin{vmatrix}
\conj{\eta}_{0, m_{1}+n}(\omega_{1}; B^{\star}) & \conj{\eta}_{0, m_{2}+n}(\omega_{1}; B^{\star})\\
\conj{\eta}_{0, m_{1}+n}(\omega_{2}; B^{\star}) & \conj{\eta}_{0, m_{2}+n}(\omega_{2}; B^{\star})
     \end{vmatrix} \\
     &\times \overlap{z_{1}, \dots, z_{N}}{m_{1}, m_{2}}_{\rm QP},
\end{split}
\end{equation}
where $\left\{\ket{m_{1}, m_{2}}_{\rm QP}\right\}$ are the states containing two CFs in the $L_{z} = m_{1}$ and $L_{z}=m_{2}$ orbitals in the $n^{\rm th}$ $\Lambda$L defined as
\begin{equation}
\begin{split}
        &\overlap{z_{1}, \dots, z_{N}}{m_{1}, m_{2}}_{\rm QP}\\
       & = \phi_{n}^{\mathrm{QP}, m_{1}, m_{2}}(z_{1}, \dots, z_{N})\times \left[\phi_{1}(z_{1}, \dots, z_{N})\right]^{2p}.
\end{split}
\end{equation}
Here, $\phi_{n}^{\mathrm{QP}, m_{1}, m_{2}}$ is a Slater determinant containing fully occupied lowest $n$ LLs plus two electrons in the $L_{z}=m_{1}$ and $L_{z}= m_{2}$ orbitals in the $n^{\rm th}$ LL. 

The QH state $\ket{\omega}_{\rm QH}$ is similarly written as
\begin{equation}\label{eq:qh-coherent-state}
\begin{split}
   & \overlap{z_{1}, \dots, z_{N}}{\omega}_{\rm QH} \\
   &\equiv \left[\sum_{m=-(n-1)}^{\infty}\eta_{0, m+(n-1)}(\omega; B^{\star})\phi^{\rm {QH}, m}_{n}(z_{1}, \dots, z_{N})\right] \\
     &\times [\phi_{1}(z_{1}, \dots, z_{N})]^{2p}\\
     & \equiv \sum_{m=-(n-1)}^{\infty}\eta_{0, m+(n-1)}(\omega; B^{\star})\Psi^{\mathrm{QH}, m}_{n}(z_{1}, \dots, z_{N})
\end{split}
\end{equation}
Again, for convenience, we shall refer to the orthogonal states with a CF absent from the $L_{z} = m$ orbital in the $(n-1)^{\rm th}$ $\Lambda$L  i.e. $\Psi^{\mathrm{QH}, m}_{n/(2pn+1)}$ as $\ket{m}_{\rm QH}$.
Note that for a finite system, we can only remove a CF from the orbitals in the $(n-1)^{\rm th}$ $\Lambda$L that are occupied and therefore, the state $\ket{\omega}_{\rm QH}$ is actually a sum of a finite number of orthogonal QH states $\left\{\ket{m}_{\rm QH}\right\}$. 
 The state containing two QHs localized at $\omega_{1}$ and $\omega_{2}$ is given by
\begin{equation}\label{eq:qh-qh-coherent-state}
\begin{split}
     &\overlap{z_{1}, \dots, z_{N}}{\omega_{1}, \omega_{2}}_{\mathrm{QH}} \\
     &= \left[\sum_{m_{1}, m_{2}=-(n-1)}^{\infty}{\eta}_{0, m_{1}+(n-1)}(\omega_{1}; B^{\star}){\eta}_{0, m_{2}+(n-1)}(\omega_{2}; B^{\star})\right.\\
     &\left.\times \phi_{n}^{\mathrm{QH}, m_{1}, m_{2}}(z_{1}, \dots, z_{N})\right]\times  [\phi_{1}(z_{1}, \dots, z_{N})]^{2p}\\
     &\equiv \sum_{m_{1} < m_{2} = -(n-1)}^{\infty}\begin{vmatrix}
{\eta}_{0, m_{1}+(n-1)}(\omega_{1}; B^{\star}) & {\eta}_{0, m_{2}+(n-1)}(\omega_{1}; B^{\star})\\
{\eta}_{0, m_{1}+(n-1)}(\omega_{2}; B^{\star}) & {\eta}_{0, m_{2}+(n-1)}(\omega_{2}; B^{\star})
     \end{vmatrix} \\
     &\times \overlap{z_{1}, \dots, z_{N}}{m_{1}, m_{2}}_{\rm QH}
\end{split}
\end{equation}
where $\left\{\ket{m_{1}, m_{2}}_{\mathrm{QH}}\right\}$ are the states with two CFs absent from the $L_{z} = m_{1}$ and $L_{z}=m_{2}$ orbitals in the $(n-1)^{\rm th}$ $\Lambda$L defined as
\begin{equation}
    \begin{split}
        &\overlap{z_{1}, \dots, z_{N}}{m_{1},m_{2}}_{\rm QH} \\
       & = \phi_{n}^{\mathrm{QH}, m_{1}, m_{2}}(z_{1}, \dots, z_{N}) 
        \times \left[\phi_{1}(z_{1}, \dots, z_{N})\right]^{2p}
    \end{split}
\end{equation}
Here, $\phi_{n}^{\mathrm{QH}, m_{1}, m_{2}}$ is a Slater determinant containing $N$ electrons filling the first $n$ LLs except for the $L_{z}=m_{1}$ and $L_{z}=m_{2}$ orbitals in the $(n-1)^{\rm th}$ LL.

We note that while the Slater determinants  $\phi_{n}^{\mathrm{QP},m}$ and $\phi_{n}^{\mathrm{QH},m}$ are properly normalized, the states obtained by composite-fermionization, that is $\Psi_{n}^{\mathrm{QP},m}$ and $\Psi_{n}^{\mathrm{QH},m}$ in Eqs.~\ref{eq:CF-qp} and \ref{eq:CF-qh}, are not normalized. This will be relevant for the analytical derivation below.

\subsection{Single QP/QH: fractional charge}

In this section, we first consider the Berry phase associated with a single QH hopping between impurities arranged at the vertices of an $M$-sided regular polygon at filling factors $\nu = 1/3$ and $\nu = 2/5$. We find that the Berry phase is equal to the AB phase acquired by a particle of charge $q=1/3$ and $q=1/5$, respectively. We then provide an analytic proof for it, which applies to all Jain fractions $\nu=n/(2pn+ 1)$ and to tunneling loops of any shape. 

\begin{figure*}[t]
\includegraphics[width=0.6\columnwidth]{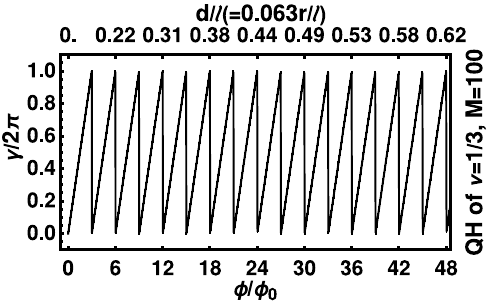}
\includegraphics[width=0.6\columnwidth]{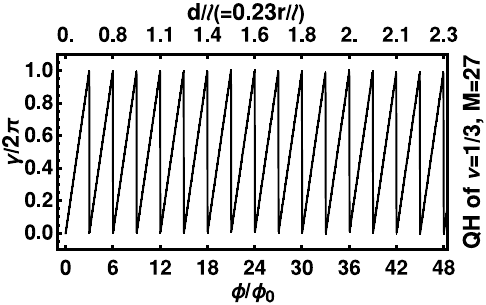}
\includegraphics[width=0.6\columnwidth]{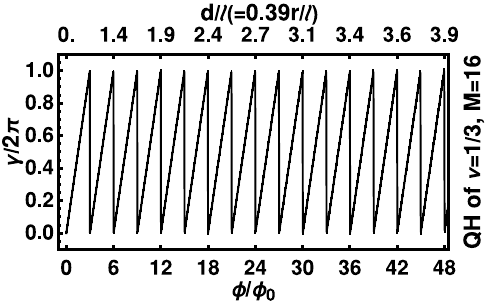}
\includegraphics[width=0.6\columnwidth]{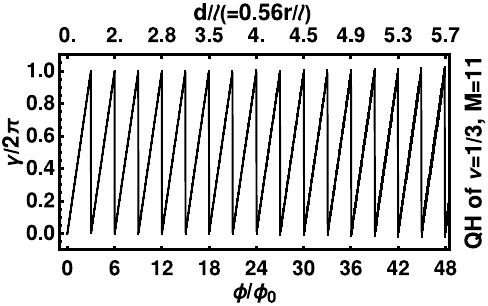}
\includegraphics[width=0.6\columnwidth]{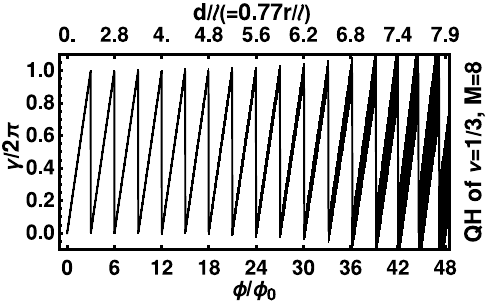}
\caption{This figure shows the numerically evaluated Berry phase ($\gamma$) associated with a closed tunnel loop of a QH at $\nu = 1/3$ around an $M$-sided regular polygon as a function of the flux $\phi$ through the polygon. Results are shown for several values of $M$ for a system of $299$ electrons. The distance $d$ separating the neighboring vertices, which is related to the distance $r$ from the center, is shown at the top of each plot for reference. The vertical width of the line in this and the subsequent figures indicates the statistical uncertainty in the Monte Carlo evaluation (arising when the tunneling amplitude becomes exponentially small). For each $M$, the Berry phase $\gamma$ is seen to be equal to $2\pi BA/\phi_0^*$ modulo $2\pi$, where $\phi_0^*=h/qe$ with $q=1/3$. As a function of the flux through the polygon, the period of oscillations is $3\phi_0$.
\label{fig:qh-1-3-charge}}
\end{figure*}

\begin{figure*}[t]
\includegraphics[width=0.6\columnwidth]{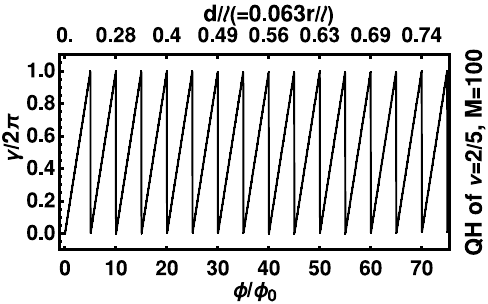}
\includegraphics[width=0.6\columnwidth]{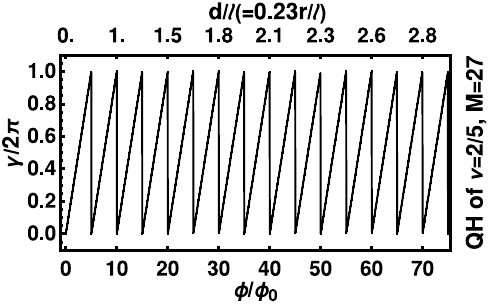}
\includegraphics[width=0.6\columnwidth]{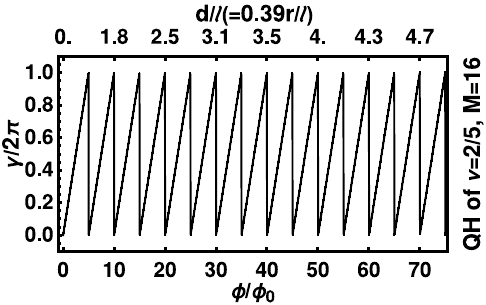}
\includegraphics[width=0.6\columnwidth]{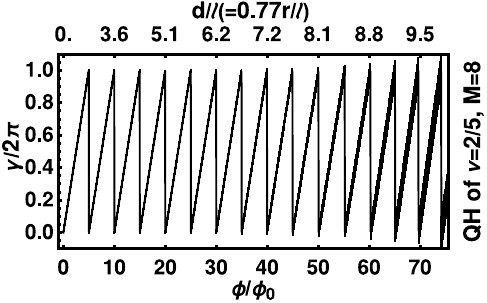}
\caption{This figure shows the numerically evaluated Berry phase ($\gamma$) associated with a closed tunnel loop of a QH at $\nu = 2/5$ around an $M$-sided regular polygon as a function of the flux $\phi$ through the polygon. Results are shown for several values of $M$ for a system of $299$ electrons. The distance $d$ separating the neighboring vertices, which is related to the distance $r$ from the center, is shown at the top of each plot for reference. For each $M$, the Berry phase $\gamma$ is seen to  be equal to $2\pi BA/\phi_0^*$ modulo $2\pi$, where $\phi_0^*=h/qe$ with $q=1/5$. As a function of the flux through the polygon, the period of oscillations is $5\phi_0$.
\label{fig:qh-2-5-charge}}
\end{figure*}

The Berry phase $\gamma$ acquired by a QH hopping along a sequence of impurities localized at $\left\{\omega_{1}, \dots, \omega_{M}\right\}$, is given by 
\begin{equation}\label{eq:berry-phase-qh}
    \gamma = \arg \left[\frac{{}_{\mathrm{QH}}\overlap{\omega_{1}}{\omega_{2}}_{\mathrm{QH}}\times {}_{\mathrm{QH}}\overlap{\omega_{2}}{\omega_{3}}_{\mathrm{QH}} \dots \times {}_{\mathrm{QH}}\overlap{\omega_{M}}{\omega_{1}}_{\mathrm{QH}}}{{}_{\mathrm{QH}}\overlap{\omega_{1}}{\omega_{1}}_{\mathrm{QH}}\times {}_{\mathrm{QH}}\overlap{\omega_{2}}{\omega_{2}}_{\mathrm{QH}} \dots \times {}_{\mathrm{QH}}\overlap{\omega_{M}}{\omega_{M}}_{\mathrm{QH}}}\right],
\end{equation}
where $|\omega\rangle_{\rm QH}$ is the state containing a QH localized at $\omega$, given in Eq.~\ref{eq:qh-coherent-state}. 
For impurities arranged at the vertices $\omega_{a} = \omega e^{\imath 2\pi a/M}$ of an $M$-sided regular polygon, the Berry phase simplifies to \begin{equation} \gamma = M\arg \left[{{}_{\mathrm{QH}}\overlap{\omega}{\omega e^{\imath 2\pi/M}}_{\mathrm{QH}}}/{{}_{\mathrm{QH}}\overlap{\omega}{\omega}_{\rm QH}}\right].
\end{equation}
We use Metropolis-Hastings-Gibbs sampling to numerically evaluate ${}_{\rm QH}\overlap{\omega}{\omega e^{\imath 2\pi/M}}_{\rm QH}/{}_{\rm QH}\overlap{\omega}{\omega}_{\rm QH}$ and thus the Berry phase $\gamma$. Our results for the Berry phase $\gamma$ (modulo $2\pi$) for a system of 299 particles at filling factors $\nu = 1/3$ and $\nu = 2/5$ are shown in Fig.~\ref{fig:qh-1-3-charge} and Fig.~\ref{fig:qh-2-5-charge} as a function of the magnetic flux enclosed by the loop for several values of $M$. The period is given by $(2pn+1)\phi_0$.

We stress here that, in our calculation of the Berry phase $\gamma$, we do not assume that the QH takes any particular path between two adjacent impurities. In other words, the microscopic calculation includes contributions from all the possible paths connecting two neighboring vertices $\omega$ and $\omega e^{\imath 2\pi/M}$. However, in each case, the calculated Berry phase $\gamma$ is equal to the AB phase of a particle of charge $q=1/(2pn+1)$ moving in the applied magnetic field $B$ along a path where the nearest neighbor impurities are connected by straight line segments.  

One can obtain this result analytically for an infinitely large system.  Here, translation invariance implies that the orthogonal states $\ket{m}_{\mathrm{QP}}$ containing a single CF in the $n^{\rm th}$ $\Lambda$L as defined in Eq.~\ref{eq:CF-qp} all have the same normalization constant, i. e., we have ${}_{\mathrm{QP}}\overlap{m}{m'}_{\mathrm{QP}} = \delta_{m,m'} \mathcal{N} (\in \mathbb{R}^{+})$ where the normalization constant $\mathcal{N}$ is independent of $m$. [This relation may be seen most readily by going to the spherical geometry (which becomes equivalent to the planar geometry in the thermodynamic limit), where these states cor-
respond to different $L_z$ states of an $L^2$ multiplet. We refer the reader to supplemental materials of Ref.~\cite{Pu22b} for explicit verification in the disc geometry.]
 Consequently, the overlap between a QP localized at $\omega_{1}$ and a QP localized at $\omega_{2}$ is given by
\begin{align*}
            &{}_{\rm QP}\overlap{\omega_{1}}{\omega_{2}}_{\rm QP} \\
            &= \sum_{m=-n}^{\infty}\eta_{0, m+n}(\omega_{1}; B^{\star})\conj{\eta}_{0, m+n}(\omega_{2};B^{\star}){}_{\rm QP}\overlap{m}{m}_{\rm QP} \\
        & = \sum_{m=-n}^{\infty}\eta_{0, m+n}(\omega_{1};B^{\star})\conj{\eta}_{0, m+n}(\omega_{2};B^{\star})\mathcal{N}\\
        &= \frac{\mathcal{N}}{2\pi}\exp\left(\imath \frac{\text{Im}\left[\conj{\omega}_{2}\omega_{1}\right]}{2\ell^{\star}{}^{2}}-\frac{\abs{\omega_{1}-\omega_{2}}^{2}}{4\ell^{\star}{}^{2}}\right) \\
\end{align*}
This implies,
\begin{equation}\label{eq:qp-qp-overlap}
    \begin{split}
        &\frac{{}_{\rm QP}\overlap{\omega_{1}}{\omega_{2}}_{\rm QP}}{\sqrt{{}_{\rm QP}\overlap{\omega_{1}}{\omega_{1}}_{\rm QP}\times {}_{\rm QP}\overlap{\omega_{2}}{\omega_{2}}_{\rm QP}}}\\
        &=\exp\left(\imath \frac{\text{Im}\left[\conj{\omega}_{2}\omega_{1}\right]}{2\ell^{\star}{}^{2}}-\frac{\abs{\omega_{1}-\omega_{2}}^{2}}{4\ell^{\star}{}^{2}}\right)
    \end{split}
\end{equation}
The phase of the overlap between states $\ket{\omega_{1}}_{\rm QP}$ and $\ket{\omega_{2}}_{\rm QP}$ can be written as
$$
    \frac{\text{Im}\left[\conj{\omega}_{2}\omega_{1}\right]}{2\ell^{\star}{}^{2}}={\hat{z} \cdot (\bm{\omega_1}\times \bm{\omega_2}) \over 2\ell^{*2}}=-2\pi {\phi\over (2pn+1)\phi_0},
$$
where $\bm{\omega}=(\omega_x,\omega_y)$ is the position vector in $\mathbb{R}^{2}$ corresponding to the complex number $\omega = \omega_{x} - \imath\omega_{y}$, and $\phi$ is the flux enclosed in the triangular area defined by $\bm{\omega}_{1}$, $\bm{\omega}_{2}$ and the origin $\bm{0}$. This can be interpreted as the AB phase $-2\pi \phi/\phi_0^*$, with $\phi_0^*=h/qe$, of a particle of charge $q=-1/(2pn+1)$. 

Similarly, the phase of the overlap between a QH localized at $\omega_{1}$ and a QH localized at $\omega_{2}$, given by
\begin{equation}\label{eq:qh-qh-overlap}
    \begin{split}
         &\frac{{}_{\rm QH}\overlap{\omega_{1}}{\omega_{2}}_{\rm QH}}{\sqrt{{}_{\rm QH}\overlap{\omega_{1}}{\omega_{1}}_{\rm QH}\times {}_{\rm QH}\overlap{\omega_{2}}{\omega_{2}}_{\rm QH}}}\\
        &=\exp\left(\imath \frac{\text{Im}\left[\conj{\omega}_{1}\omega_{2}\right]}{2\ell^{\star}{}^{2}}-\frac{\abs{\omega_{1}-\omega_{2}}^{2}}{4\ell^{\star}{}^{2}}\right),
    \end{split}
\end{equation}
can be interpreted as the AB phase of a particle of charge $q=1/(2pn+1)$.

This proves that the Berry phase $\gamma$ for {\it any} closed tunnel loop of a QP/QH is equal to the AB phase of a particle of charge $q=\mp 1/(2pn+1)$ going around that loop. 

\subsection{Two QHs: Fractional statistics}

We next consider the change in the Berry phase associated with a closed tunneling loop when another QH is added in the interior. For this purpose, it is sufficient to work with a regular polygon path with the additional QH at its center. As discussed below, we find that the additional Berry phase (i.e. braiding phase) is equal to the expected value of $\theta_{\mathrm{QH}-\mathrm{QH}} = 4\pi/3$ at $\nu = 1/3$ and $\theta_{\mathrm{QH}-\mathrm{QH}} = 4\pi / 5$ at $\nu = 2/5$, provided that the polygon encloses the central QH in its entirety. 

\begin{figure*}[t]
\includegraphics[width=0.6\columnwidth]{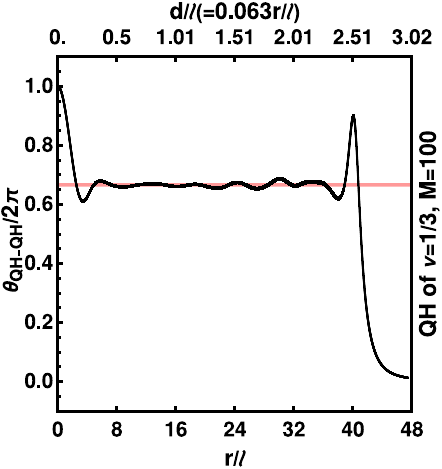}
\includegraphics[width=0.6\columnwidth]{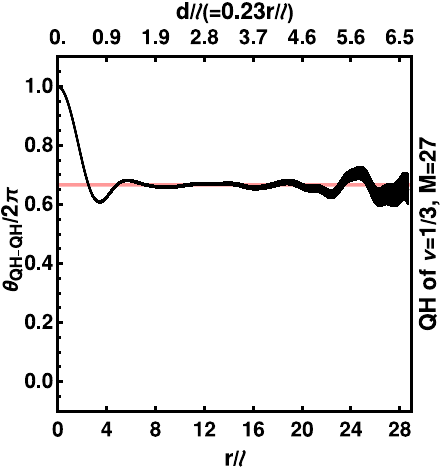}
\includegraphics[width=0.6\columnwidth]{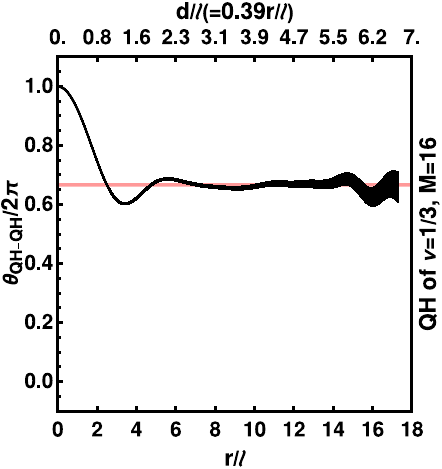}
\includegraphics[width=0.6\columnwidth]{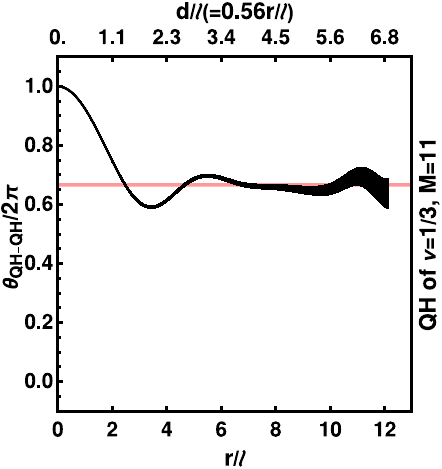}
\includegraphics[width=0.6\columnwidth]{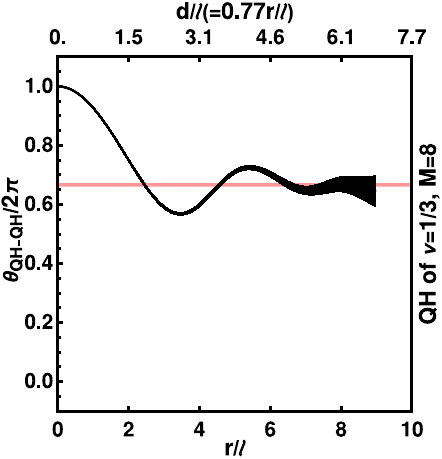}
\caption{This figure shows the QH statistics $\theta_{\mathrm{QH}-\mathrm{QH}}$ for $\nu=1/3$, which is proportional to the change in the Berry phase accrued by a QH along a closed tunneling loop 
when another QH is inserted at the center. The loop considered here is an $M$-sided regular polygon of radius $r$; $\theta_{\mathrm{QH}-\mathrm{QH}}$ is depicted as a function of $r$ for several values of $M$. The system contains $298$ electrons at $\nu=1/3$. In each case, we have shown only the region (i.e. range of $r$) in which the statistical error in $\theta_{\mathrm{QH}-\mathrm{QH}}$ arising from Monte Carlo sampling (indicated by the line width) is less than $2\pi/10$. [The Monte Carlo error becomes more significant as the strength of overlap ($=e^{-d^2/12\ell^{2}}$, see Eq.~\ref{eq:qh-qh-overlap}) between the QH wave functions centered at two successive vertices decreases.] Within Monte Carlo error, $\theta_{\mathrm{QH}-\mathrm{QH}}$ takes on the expected value $4\pi/3$ for all $M$ when the tunneling path completely encloses the central QH, i.e. for $r \geq 7 \ell$. The deviation from the $4\pi/3$ at large $r$ is due to proximity to the edge of the FQH droplet.
\label{fig:qh-1-3-statistics}}
\end{figure*}

\begin{figure*}[t]
\includegraphics[width=0.6\columnwidth]{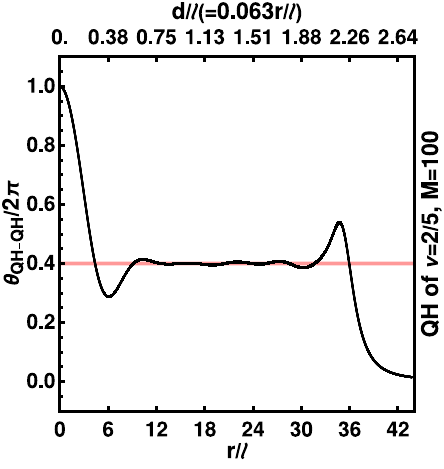}
\includegraphics[width=0.6\columnwidth]{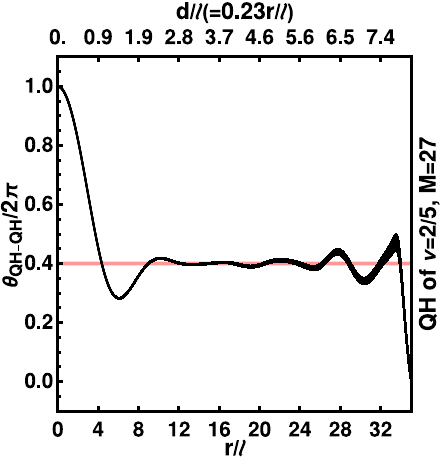}
\includegraphics[width=0.6\columnwidth]{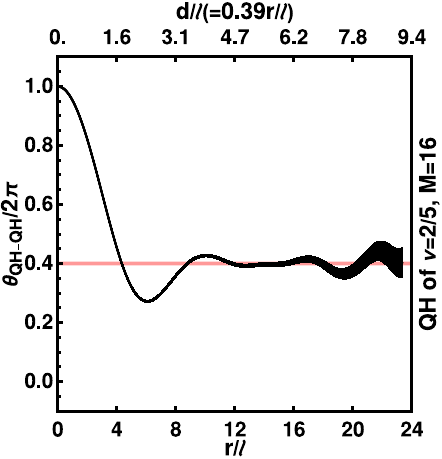}
\includegraphics[width=0.6\columnwidth]{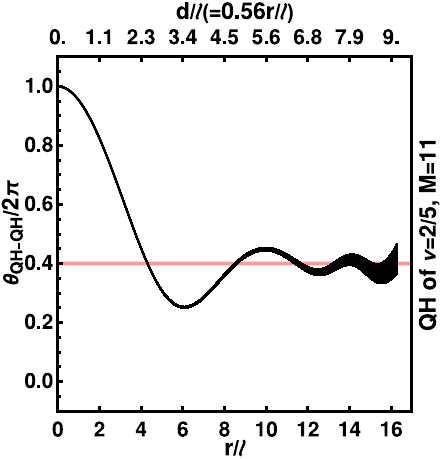}
\includegraphics[width=0.6\columnwidth]{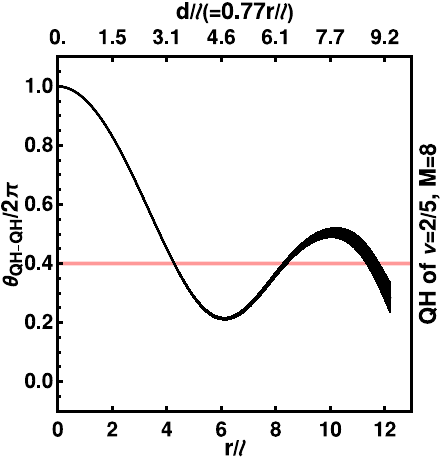}
\caption{Same as Fig.~\ref{fig:qh-1-3-statistics} but for the QHs of $\nu=2/5$. 
Within Monte Carlo error, $\theta_{\mathrm{QH}-\mathrm{QH}}$ takes on the expected value $4\pi/5$ for all $M$ when the tunneling path completely encloses the central QH, i.e. for $r \geq 12 \ell$ and is far from the edge. 
\label{fig:qh-2-5-statistics}}
\end{figure*}

We place a QH at the origin and consider another QH hopping along the vertices $\omega_{a} = \omega e^{\imath 2\pi a/M}$ of an $M$-sided regular polygon. We  numerically evaluate ${}_{\rm QH}\overlap{0, \omega}{0, \omega_{a}}_{\mathrm{QH}}/{}_{\mathrm{QH}}\overlap{0, \omega}{0, \omega}_{\rm QH}$ (here, $\ket{\omega_{1}, \omega_{2}}_{\rm QH}$ is the state containing two QHs localized at $\omega_{1}$ and $\omega_{2}$; see Eq.~\ref{eq:qh-qh-coherent-state}) which is related to the Berry phase $\gamma$ associated with the full tunnel loop as
$$
\gamma = M\arg\left[\frac{{}_{\rm QH}\overlap{0, \omega}{0, \omega e^{\imath 2\pi/M}}_{\rm QH}}{{}_{\rm QH}\overlap{0, \omega}{0, \omega}_{\rm QH}}\right]
$$
Subtracting from it the Berry phase 
obtained in the previous section yields $\theta_{\mathrm{QH}-\mathrm{QH}}$ (modulo $2\pi$), which we present 
for a system of 298 particles at filling factors $\nu = 1/3$ and $\nu = 2/5$ as a function of $r = \abs{\omega}$ in Fig.~\ref{fig:qh-1-3-statistics} and Fig.~\ref{fig:qh-2-5-statistics} for several values of $M$.

For sufficiently large $r = \abs{\omega}$, $\theta_{\mathrm{QH}-\mathrm{QH}}$ takes on a constant value, within Monte Carlo error, equal to $4\pi/3$ at $\nu = 1/3$ and $4\pi/5$ at $\nu = 2/5$ for all $M$. There are corrections to this value when the closed tunneling loop does not fully contain the central QH, i.e. when $r$ is less than the radius of the QH. For $\nu = 1/3$, the radius of the QH is $r \approx 7 \ell$, and for $\nu = 2/5$, the radius is $r \approx 12 \ell$ (see Fig.~\ref{fig:qh-densities}). Finally, as noted in the introductory section, there is a correction also when the tunnelling loop approaches the edge of the system. (For a system of $298$ particles, the radius of the $\nu = 1/3$ droplet is $\sim 43\ell$ and the radius of the $\nu = 2/5$ droplet is $\sim 39 \ell$.)

\subsection{Remarks}

The above Berry phase associated with the tunneling loop implies periodic oscillations in the tunnel conductance, which is expected to show a maximum (minimum) when the paths along the upper half and the lower half of the tunneling loop interfere constructively (destructively). The period of oscillations of the tunnel conductance as a function of the flux $\phi$ for the FQH state at $\nu=n/(2pn+1)$ is given by $(2pn+1)\phi_0$, which, interpreted as $h/(qe)$, implies a fractional charge $q=1/(2pn+1)$.  
The addition of another QP/QH in the interior of the loop will cause a shift in the oscillations (see Eq.~\ref{eq:CF-qp-qp-qp-statistics},~Eq.~\ref{eq:CF-qh-qp-qh-statistics} and Eq.~\ref{eq:qh-qh-statistics}) that is related to the fractional braid statistics.

We have considered above only the  braid statistics for QHs, namely $\theta_{\rm QH-QH}$. Noting that a QP-QH pair is a boson, it follows that $\theta_{\rm QH-QP}=-\theta_{\rm QH-QH}$, and similarly $\theta_{\rm QP-QP}=\theta_{\rm QH-QH}$. The braid statistics  $\theta_{\rm QP-QP}$ and  $\theta_{\rm QP-QH}$ can also be evaluated from explicit calculations using the wave functions~\cite{Jeon03b,Jeon04}.
It is worth recalling that the calculation of $\theta_{\rm QP-QP}$ involves a subtlety, namely that the real distance between two QPs, which are excited CFs, is different from the `apparent' distance due to the repulsion introduced by vortex attachment~\cite{Jeon03b,Jeon04}. It is necessary to carefully account for that effect to determine the correct braid statistics~\cite{Jeon03b,Jeon04}. Above we have worked with QHs where this is not an issue, because no vortices are attached to {\it missing} CFs.

\section{Discussion}
\label{sec:Discussion}

In this article, we begin by noting that fractional charge and fractional statistics are not sharply defined at the edges of a FQH system because of the absence of a gap there. We argue that the interference experiment of Ref.~\cite{Nakamura20} can be understood only by assuming that CFs are well defined at the edge. 

We then propose a possible scheme for measuring the braid statistics of the QPs or QHs in a manner such that the object moving in a closed loop has a fractional charge over the entire loop. The idea is to create a tunneling loop for a QP/QH in the bulk of the sample by placing a sequence of impurities and then considering transport through it, as shown in Fig.~\ref{fig:setup}. We show that the conductance for  a given tunneling path displays periodic oscillations as a function of the magnetic flux through it, with the period given by $(2pn+1)\phi_0$ for FQH state at $\nu=n/(2pn+1)$, which, interpreted as $h/qe$ implies a fractional charge $q=1/(2pn+1)$. Furthermore, adding another QP/QH in the interior of the loop causes a shift in the oscillations that is related to the fractional braid statistics. 

The measurement of fractional statistics with this geometry will require several conditions which we list. (i) The conductance should be determined predominantly by a single loop with a well defined area. That is the case when only hopping between the nearest neighbor impurities is relevant, i.e.,  the distance between the nearest neighbor impurities is large compared to the QP/QH size (so the tunneling into next to nearest neighbor impurity is negligible). Of course, the distance between the nearest neighbor impurities ought not to be too large either, because then the signal would be very small and also variations in the impurity potential will tend to localize the QP/QH on a single impurity. (In Appendix~\ref{app:1} we evaluate the energy of a tunneling hole at $\nu = 1$ without assuming only nearest neighbor tunneling to demonstrate that oscillations with a well defined period occur only when the nearest neighbor tunneling dominates.)  (ii) The QP/QH added in the interior should be contained entirely within the tunneling loop. (iii) The addition of a QP/QH in the interior of the loop should not change the area of the tunneling loop. This can be arranged by screening the potential of the QP/QH by nearby metallic gates or by making the impurity potential sufficiently sharp. (iv) The impurities should roughly have the same local potential, that is, they should be as identical as possible for the QP/QH to resonantly tunnel between nearby impurities.

J.K.J. is grateful to Steven Kivelson for insightful remarks and conversations. M.G. and J.K.J. acknowledge financial support from the U.S. National Science Foundation under Grant No. DMR-2037990.  Computations for this research were performed on the Pennsylvania State University's Institute for Computational and Data Sciences' ROAR supercomputer.

\begin{appendices}
\section{}\label{app:1}
We explicitly evaluate below the energy of a tunneling hole at $\nu = 1$ and use it as an example to illustrate that there is indeed a regime where the tunneling path (and thus the area enclosed by the path) becomes well-defined, leading to oscillations with a well-defined period.
Let us consider $\delta$-function impurities placed at the $M$ vertices $\omega_{a} = \omega e^{\imath 2\pi a/M}$ such that $a \in \left\{1, \dots, M\right\}$ of an $M$-sided regular polygon. The impurities produce a potential $\sum_{a=1}^{M}\sum_{i}\delta(z_{i}-\omega_a)$. Let us denote a hole localized at $\omega_{a}$ by $\ket{\omega_{a}}$, as defined in Eq.~\ref{eq:qh-coherent-state}, 
A hole in the presence of $M$ impurities must exist in a linear superposition of states $\ket{\omega_{a}}$. 
From the $M$-fold rotational symmetry of the potential $\sum_{a=1}^{M}\sum_{i}\delta(z_{i}-\omega_{a})$, it follows that the (unnormalized) eigenstates are
\begin{equation}
 \ket{k} = \sum_{a=1}^{M}e^{\imath k a}\ket{\omega_{a}}
\end{equation}
with $k \in \left\{0, \frac{2\pi}{M}, \dots, \frac{2\pi (M-1)}{M}\right\}$. Therefore, the energy of the hole is given by
\begin{equation}\label{eq:bloch-energy}
 E_{k} = \frac{\bra{k}\left(\sum_{a=1}^{M}\sum_{i}\delta(z_{i}-\omega_{a})\right)\ket{k}}{\overlap{k}{k}}
\end{equation}
In Ref.~\cite{Pu22b}, it is shown that for an infinitely large system,
\begin{widetext}
    \begin{equation}\label{eq:iqhe-formulae}
        \begin{split}
            &\frac{\overlap{\omega_{1}}{\omega_{2}}}{\sqrt{\overlap{\omega_{1}}{\omega_{1}}\overlap{\omega_{2}}{\omega_{2}}}} = \exp\left(\imath \frac{\text{Im}\left[\conj{\omega}_{1}\omega_{2}\right]}{2\ell^{2}}-\frac{\abs{\omega_{1}-\omega_{2}}^{2}}{4\ell^{2}}\right) \\
            &\frac{\bra{\omega_{1}}\left(\sum_{i}\delta(z_{i}-\omega)\right)\ket{\omega_{2}}}{\sqrt{\overlap{\omega_{1}}{\omega_{1}}\overlap{\omega_{2}}{\omega_{2}}}} = \frac{1}{2\pi}\frac{\overlap{\omega_{1}}{\omega_{2}}}{\sqrt{\overlap{\omega_{1}}{\omega_{1}}\overlap{\omega_{2}}{\omega_{2}}}}- \frac{1}{2\pi}\frac{\overlap{\omega_{1}}{\omega}}{\sqrt{\overlap{\omega_{1}}{\omega_{1}}\overlap{\omega}{\omega}}}\frac{\overlap{\omega}{\omega_{2}}}{\sqrt{\overlap{\omega}{\omega}\overlap{\omega_{2}}{\omega_{2}}}}
        \end{split}
    \end{equation}    
\end{widetext}
Substituting Eq.~\ref{eq:iqhe-formulae} in Eq.~\ref{eq:bloch-energy}, we find the following enlightening form for the energy of the hole:
\begin{widetext}
    \begin{equation}\label{eq:bloch-energy-final}
 E_{k} = \frac{M-1}{2\pi} - \frac{1}{\pi}\sum_{a=1}^{\lfloor \frac{M-1}{2} \rfloor}\exp \left(-\frac{\omega^{2}}{\ell^{2}}\sin^{2}(\frac{\pi a}{M})\right)\cos (ka + \frac{\omega^{2}}{2\ell^{2}}\sin(\frac{2\pi a}{M})) - \delta(M \equiv 0 \; \mathrm{mod} \; 2)\frac{1}{2\pi}\exp \left( -\frac{\omega^{2}}{\ell^{2}} + \frac{\imath k M}{2}\right)
\end{equation}
\end{widetext}
The above equation has a very intuitive interpretation:
A hole tunneling through a setup such as the one shown in Fig.~\eqref{fig:setup}, can take any one of the $\lfloor (M-1) / 2\rfloor$ tunneling paths each enclosing a different area. The tunneling paths involve the hole moving along straight lines connecting neighboring impurities, next nearest impurities and so on. 
However, the paths are not equiprobable. The Gaussian shape of the hole wavepacket implies that a tunneling path connecting impurities $a$ sites away (with the nearest impurity being $1$ site away) is weighted by $e^{-d^{2}/4\ell^{2}}$ where $d = 2\omega \sin (\pi a / M)$ is the distance between impurities along the path. This is crucial as it allows us to tune the distance between impurities such that the tunneling path connecting neighboring impurities contributes significantly whereas the contributions from the other paths are heavily suppressed. Under such conditions, Eq.~\ref{eq:bloch-energy-final} tells us that the energy oscillates as a function of the flux enclosed ($=M\phi_{0}\omega^{2}\sin(2\pi/M)/4\pi\ell^{2}$) with a period equal to $\phi_{0}$. 

\end{appendices}

%
\end{document}